# A Universal Method to Generate Hyperpolarisation in Beams and Samples

R. Engels[1,2,†], T. El-Kordy[1,2,3], N. Faatz[1,2,4], C. Hanhart[1,5], N. Hanold[1,6], C.S. Kannis[1,2], L. Kunkel[1,2], S. Pütz[1,2], H. Sharma[1,2,3], T. Sefzick[1,2], H. Soltner[7], V. Verhoeven[1,2], M. Westphal[1,2,3], J. Wirtz[1,2,4] and M. Büscher[6,8]

[1] Institut für Kernphysik, Forschungszentrum Jülich, Jülich, Germany

[2] GSI, Helmholtzzentrum für Schwerionenforschung, Darmstadt, Germany

[3] FH Aachen, Campus Jülich, Jülich, Germany

[4] III. Physikalisches Institut B, RWTH Aachen University, Aachen, Germany

[5] Institute for Advanced Simulation 4, Forschungszentrum Jülich, Jülich, Germany

[6] Institut für Laser- und Plasma-Physik, Heinrich-Heine-Universität Düsseldorf, Düsseldorf, Germany

[7] Zentralinstitut für Engineering, Elektronik und Analytik, Forschungszentrum Jülich, Jülich, Germany

[8] Peter-Grünberg-Institut 6, Forschungszentrum Jülich, Jülich, Germany

[†] e-mail: r.w.engels@fz-juelich.de

**Summary**

Sizable hyperpolarisation, i.e. an imbalance of the occupation numbers of nuclear spins in a sample deviating from thermal equilibrium, is needed in various fields of science. For example, hyperpolarised tracers are utilised in magnetic resonance imaging in medicine (MRI) and polarised beams and targets are employed in nuclear physics to study the spin dependence of nuclear forces. Here we show that the quantum interference of transitions induced by radio-wave pumping with longitudinal and radial pulses are able to produce large polarisations at small magnetic fields. This method is easier than established methods, theoretically understood and experimentally proven for beams of metastable hydrogen atoms in the keV energy range. It should also work for a variety of samples at rest. Thus, this technique opens the door for a new generation of polarised tracers, possibly low-field MRI with better spatial resolution or the production of polarised fuel to increase the efficiency of fusion reactors by manipulating the involved cross sections.

For decades, great efforts have been made to hyperpolarise atoms, molecules or even solids, i.e. to align the nuclear spins in a preferred direction, which is then much more occupied than would be expected in thermal equilibrium. For example, the polarisation induced by superconducting solenoids for magnetic-resonance imaging (MRI) in medicine is on a level of $P=(N_\uparrow - N_\downarrow)/(N_\uparrow + N_\downarrow) \sim 10^{-5}$, but for hyperpolarised tracers values of $P\sim0.5$ are reached to increase the resolution of MRI by orders of magnitude [1,2]. Moreover, polarised beams and targets are employed in nuclear physics to investigate the spin-dependence of the nuclear forces [3] and, based on this, it was proposed already 90 years ago [4] to use these effects to increase the efficiency of fusion reactors in different ways (see Ref. [5] for a recent discussion), but to date no techniques are available to produce the necessary amounts of highly polarised fuel.

In recent years, multiple methods have been established to produce hyperpolarisation. The *brute force method* uses strong magnetic fields (~15 T) and low temperatures (~25 mK) to build up nuclear polarisation for HD molecules on a typical time scale of many hours [6]. Polarised photons of dedicated lasers can induce *optical pumping* of the electron spin in alkali atoms that is transferred to nucleons by hyperfine interactions [7,8]. *Dynamic-nuclear polarisation* employs microwave pumping of electrons in radicals at low temperatures (1–100 K) and strong magnetic fields (1–3 T) [9]. In polarised atomic beam sources (ABS) a combination of Stern-Gerlach magnets and radio-frequency induced transitions between single hyperfine substates is used for *spin filtering* of atomic hydrogen or deuterium beams [10]. All these methods are either very expensive, limited in terms of polarisation levels and intensity, or restricted to single atomic isotopes or dedicated molecules.

In the following a universal and cheap method to polarise atoms, molecules and their ions is introduced. This technique employs a coherent and monochromatic single radio-wave pulse where the corresponding photons induce magnetic dipole transitions between Zeeman substates at low magnetic field, which then interfere with each other. By that, polarisation values between $0.1 < P < 0.9$ are possible at room temperatures even for macroscopic probes of many different materials.

The interaction of an atom with an external magnetic field $B(t)$ is described by the corresponding Hamilton operator

$$\begin{aligned} H(t) &= \Delta E_{\text{HFS}} \, \boldsymbol{I} \cdot \boldsymbol{J} - \boldsymbol{\mu}_{\text{atom}} \cdot \boldsymbol{B}(t) \\ &= \Delta E_{\text{HFS}} \, \boldsymbol{I} \cdot \boldsymbol{J} - (-g_J \mu_B \boldsymbol{J} + g_p \mu_N \boldsymbol{I}) \cdot \boldsymbol{B}(t) \\ &= H_0 + V(t) \quad , \end{aligned}$$

where $H_0$ denotes the hyperfine splitting itself, a time-independent term with the hyperfine splitting energy $\Delta E_{\text{HFS}}$. Here $g_J = 2.001$ and $g_p = 5.586$ are the g-factors of the electron and the proton, respectively, and $\mu_B$ and $\mu_N$ are the electron and nuclear magneton. The only tunable quantity is the magnetic field $B(t)$. In particular, the interaction with an oscillating field $B(t)$ of an induced electromagnetic wave was discussed in Refs. [11, 12] and is used by an ABS to induce transitions between single hyperfine substates (HFS).

As demonstrated in Ref. [13] there is an alternative option for generating an electromagnetic wave: Dedicated counter-rotating coils with the same number of windings and the same current produce a longitudinal sinusoidal magnetic field with a wavelength $\lambda$ (see Fig. 1). When hydrogen atoms with a given velocity $v$ at a radial distance $r$ pass through this static magnetic field $B_z(z)$, the atoms experience an incoming single electromagnetic pulse in their rest frame with a time dependent magnetic field amplitude $B(t)$. Of course, the velocity of this incoming wave $v$ is much smaller than the speed of light $c$.

In the laboratory system Gauss' law for magnetism $\vec{\nabla} \cdot \vec{B} = 0$ can be used to calculate the radial field component $B_r$ from the given longitudinal $B_z$ to

$$B_r(r,z) = -\frac{r}{2}\frac{\partial B_z(z)}{\partial z},$$

which is also directly measurable (see Fig. 1).

While the longitudinal magnetic field component induces the Zeeman splitting and is responsible for the hyperfine beat [14] between the substates with $m_F=0$, the corresponding radial field leads to transitions with $\Delta m_F = \pm 1$. The calculations are performed in the rest frame of the hydrogen atom where the static magnetic field acquires a time dependence. Using time-dependent perturbation in $V(t)$ the Schrödinger equation can be transformed into a set of coupled differential equations

$$i\hbar \frac{\partial c_k(t)}{\partial t} = \sum_{i=1}^{4} c_i(t) e^{-i(E_i - E_k)t/\hbar} \langle k|V(t)|n\rangle,$$

where $c_i(t)$ describes the amplitude of the i-th state and the sum runs over all relevant states. In case of the metastable 2S states, which are used here to demonstrate the method, the 2P states can be neglected, since each set of these states has a large energy separation of two to three orders of magnitude at magnetic fields below 10 mT. The solutions of the equations are averaged over a Gaussian beam profile in $r$ as well as a velocity profile. In addition, the occupation numbers of the HFS at the beginning and magnetic field inhomogeneities, especially that $B_z(r)$ is not constant, are taken into account.

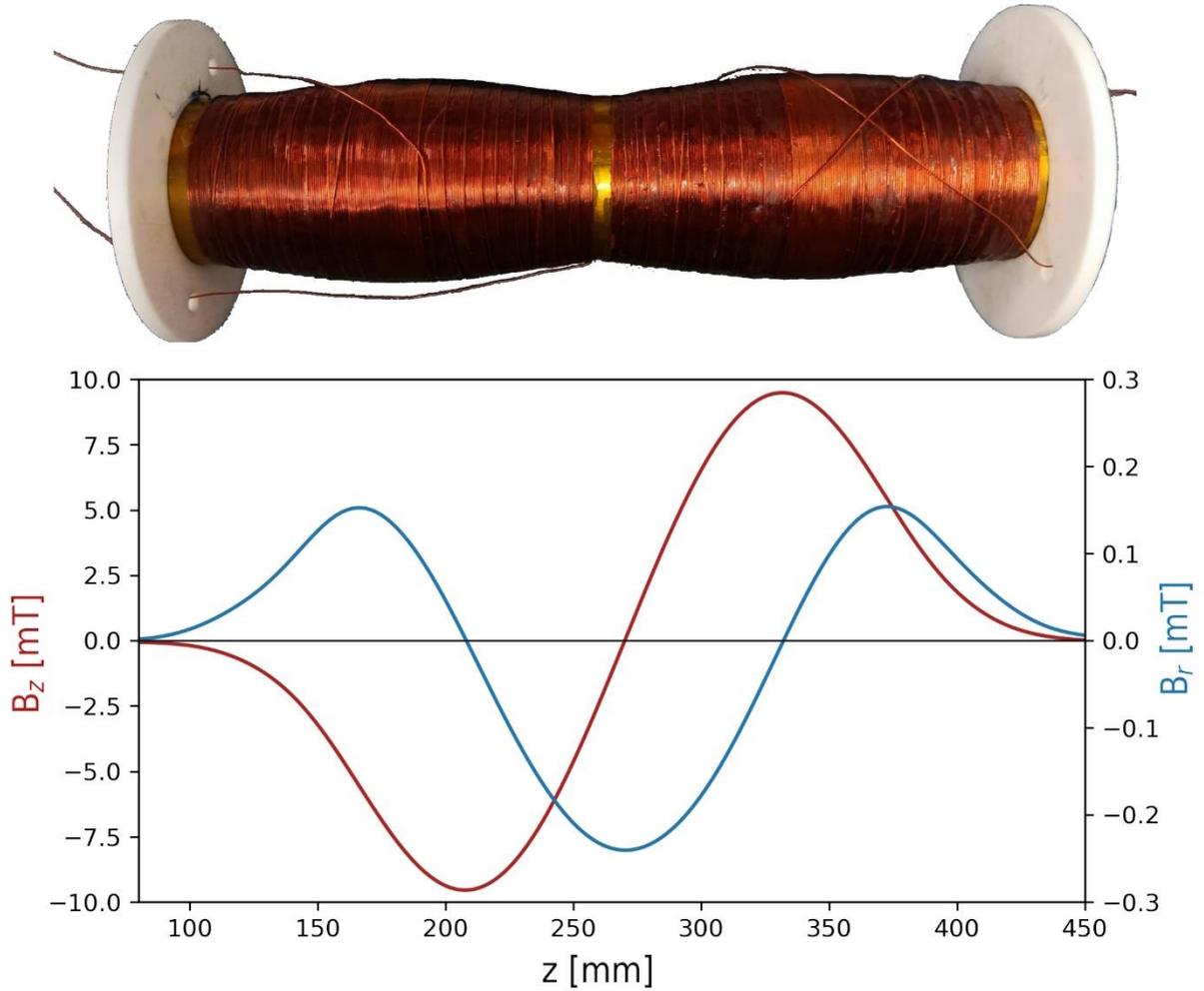

**Figure 1:** The two opposing coils that are designed to induce a sinusoidal magnetic field and the measured longitudinal (red) and the corresponding radial (blue) fields at a distance of $r=5$ mm from the symmetry axis for a coil current of 0.5 A in the coils.

When hydrogen atoms in certain HFS pass through such a field configuration, the occupation numbers start to oscillate within the substates of the $F=1$ multiplet. This behaviour was observed before for (metastable) hydrogen atoms [15,16], but could not be described successfully [17,18]. In Ref. [13] we already published results of measurements and simulations for metastable beams in HFS α1 passing this configuration. What is new in this work is the study of an unpolarised beam of metastable hydrogen, i.e. all four HFS equally populated ($|c_i|^2=0.25$), sent through such a magnetic field configuration. This setting reveals similar oscillations of the occupation numbers (Fig. 2) which opens great opportunities to polarise a huge class of materials.

In the current setting a beam of metastable hydrogen atoms in different HFS can be produced with components of a Lamb-shift polarimeter in front of the static oscillating magnetic field. Afterwards, other parts are used to determine the occupation numbers of single HFS with the electron spin $m_J=+1/2$, i.e. the α substates, as function of the current sent through the coils. The measured spectra for the different starting conditions $|c_{\alpha 1}|^2=1$ and $|c_i|^2\approx 0.25$, i.e. a nearly unpolarised beam, are shown in Fig. 2.

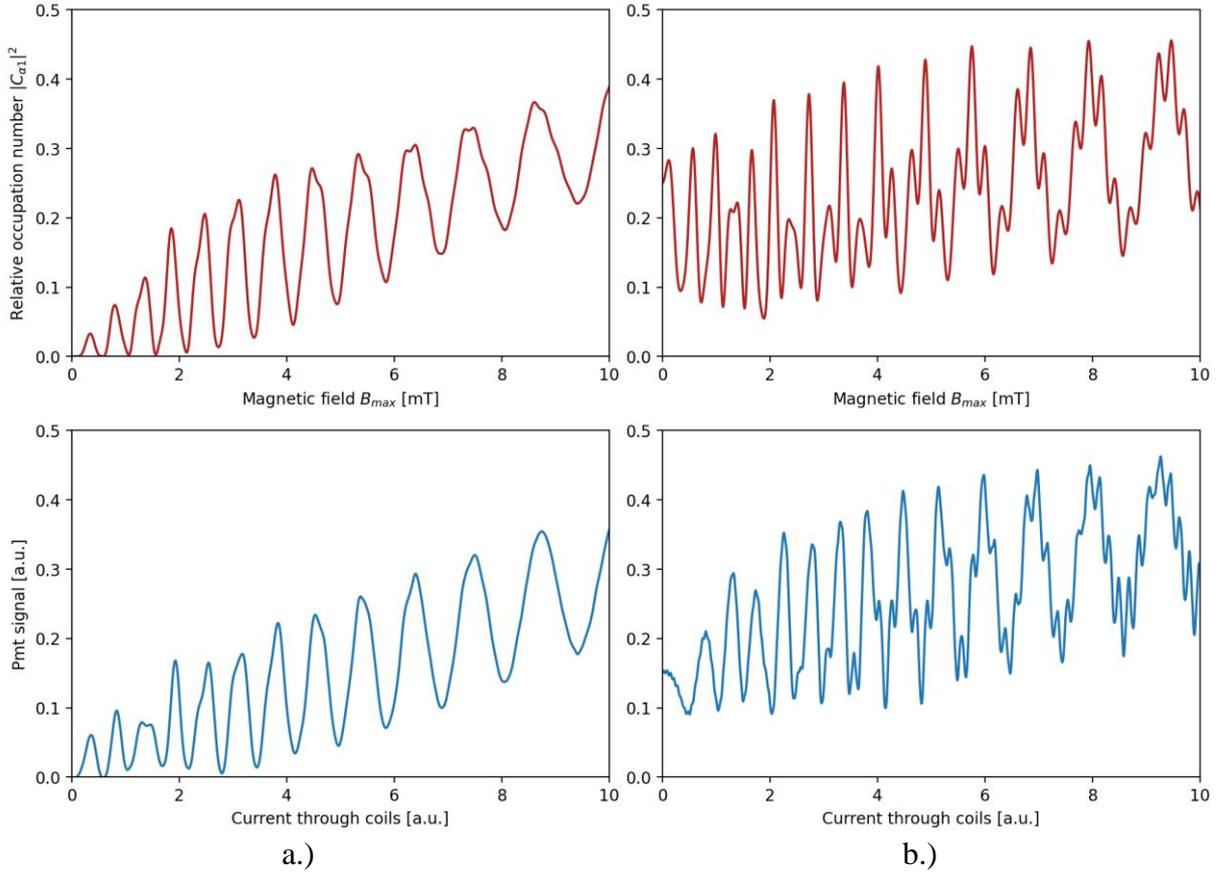

**Figure 2:** Calculated (red) and measured (blue) occupation numbers of atoms in the HFS α1 behind the coils when a beam of metastable hydrogen atoms at 1.5 keV passes through the magnetic field from Fig. 1 with $|c_{\alpha 1}|^2=1$, all atoms are in HFS α1 (a.), and $|c_i|^2 \approx 0.25$, i.e. a nearly unpolarised beam (b.). The x- and y-scale of the measured spectra are calibrated due to the simulations.

A measurement with a completely unpolarised metastable hydrogen beam is not possible with our setup, because a longitudinal magnetic field in front of the opposing coils is needed to define the quantisation axis of the total spin $F$ along the beam axis. In the additional presence of electric fields, the lifetime of the single hyperfine states is changed. These electric components are created by the radial magnetic part in the rest frame of the beam and, therefore, the lifetime of the α substates ($m_J=+1/2$) becomes slightly larger than that of the β substates ($m_J=-1/2$) [19]. This leads to a non-zero electron polarisation as function of this longitudinal magnetic field and, due to different starting conditions, to slightly different measured spectra than shown in Fig. 2b.). Other effects influencing the simulations are the velocity distribution of the beam and its profile, which slightly broadens the resonances. In addition, magnetic field inhomogeneities are increasing the half width and change the magnetic field calibration.

The observed transitions can be understood straightforwardly: During their time-of-flight $\Delta t$ through the sinusoidal magnetic field the atoms experience a magnetic field $B$ that is the averaged square of the absolute value of the magnetic field amplitude $B_z(z,r)$. If the magnetic field oscillation were a perfect sine function, this magnetic field would be $B = B_{max}/\sqrt{2}$, with

$B_{max}$ being the maximum magnetic field amplitude in the centre of a coil. This magnetic field $B$ defines the energy splitting of the HFS in the Breit-Rabi diagram of Fig. 3.

In the rest system of the atoms the static magnetic field oscillation with wavelength $\lambda$ appears as an electromagnetic wave with frequency

$$f_0 = 1/\Delta t = v/\lambda \quad .$$

The radial oscillation of the static magnetic field, which has the same $\lambda$, can be interpreted as a mono-energetic and coherent single photon pulse with an energy of

$$E_{Ph} = h \cdot f_0 = h \cdot v/\lambda \quad .$$

Thus, the energy of these photons can be tuned either by the beam energy that defines the beam velocity $v$ or by the distance between the coils that defines the wavelength.

For a beam of metastable hydrogen atoms with an energy of 1.5 keV ($v \sim 6.1 \cdot 10^5$ m/s) and $\lambda \sim 0.25$ m the frequency is $f_0 \sim 2.45$ MHz and the photon energy is only $E_{Ph} \sim 10$ neV. These photons can induce magnetic dipole transitions with $\Delta m_F = \pm 1$ within the hyperfine splitting energies of the $F=1$ multiplet every time the energy difference between the substates is

$$E(B) = (2n-1) \cdot f_0 \cdot h \quad ,$$

where $n$ is an integer. Note that only the absorption of an odd number of photons is allowed due to angular–momentum conservation. Therefore, at different magnetic field amplitudes inside the coils, resonant enhancements of the occupation numbers of the single substates are observed (see Fig. 3).

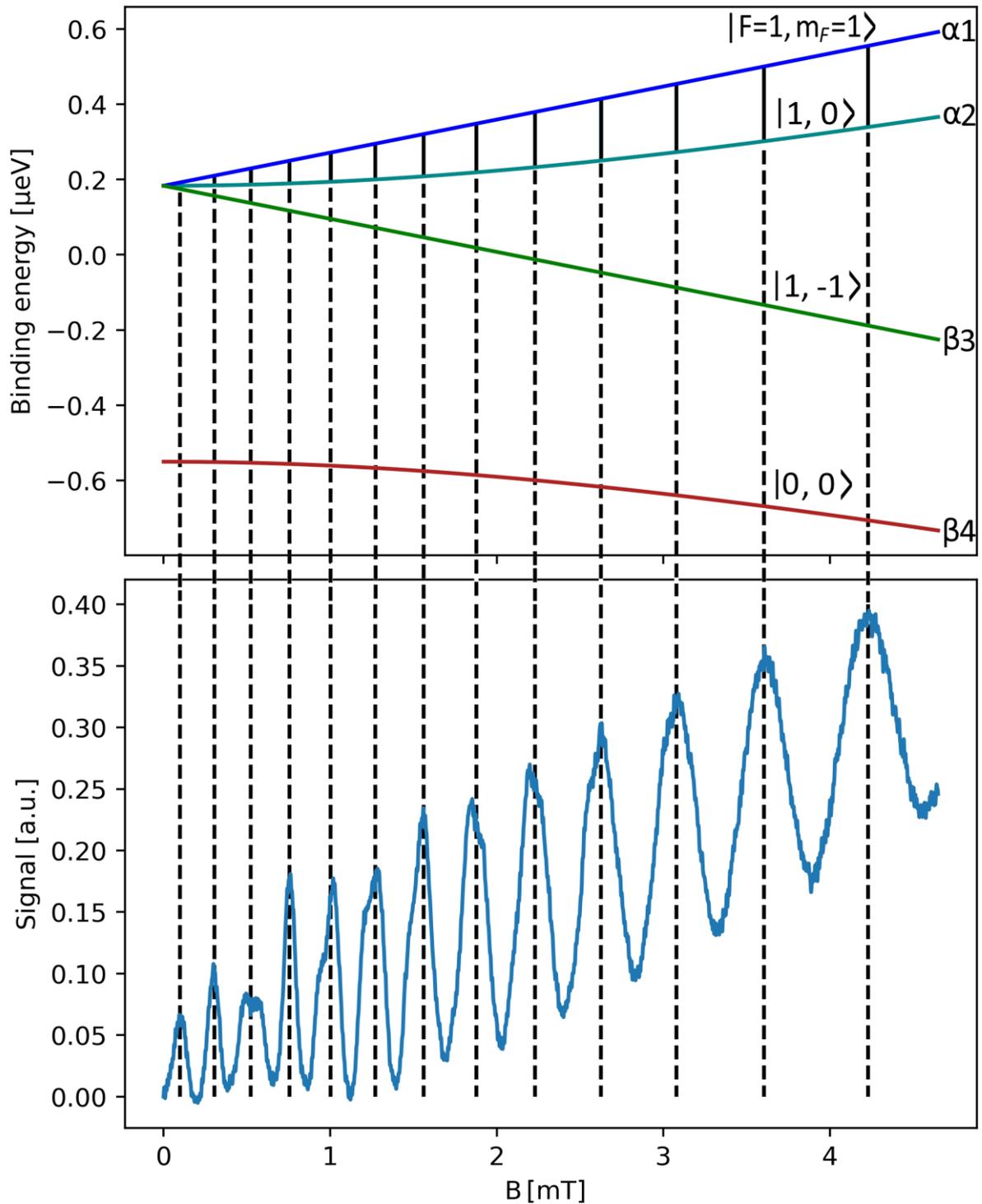

**Figure 3:** Upper panel: Breit-Rabi diagram, i.e. the binding energy as a function of an external magnetic field, of hydrogen atoms in the metastable $2S_{1/2}$ state. Lower panel: Occupation number of the α1 level as a function of the magnetic field, when initially only α1 is occupied. Naively the Sona transition empties this level, however, every time the energy gap between α1 and α2 is an odd multiple of the photon energy $E_{Ph}$ ~10 neV a transition between these states is possible and the HFS α1 can be populated again. In addition, the transition α2↔β3 deforms the shape of the resonance, because some atoms can be lost from α2 into β3.

As shown in Fig. 4 these transitions can be induced mostly before and after the zero-crossing between the magnets, because the radial component shows 1.5 oscillations inside the coils. Nevertheless, both transitions must share one half of the magnetic field oscillation. In addition, a so-called Sona transition, i.e. an exchange of the occupation numbers between the substates *α1 ($m_F=+1$)* and *β3 ($m_F=-1$)* [20], takes place at the zero crossing of the magnetic field $B_z$ between the coils. Consequently, the direction of the quantisation axis, namely the magnetic field direction, is exchanged faster than the spins could follow due to their Larmor precession. Since the beam is relatively fast and the small radial field component induces a slow Larmor precession, this kind of non-adiabatic transition is almost perfect.

For example, if a beam of metastable atoms in HFS α1 only enters the magnetic field, the atoms end up in β3 behind the solenoids when no photons are absorbed. If the α1↔α2 transition is induced by the photons, the atoms can decay into α2, survive the zero-crossing in this state and then absorb another photon to populate α1 again. When the α2↔β3 transition is excited in parallel, some atoms may change from state α2→β3 and cannot return to HFS α1 anymore. Thus, the resonance from α2→α1 is deformed by the losses into β3. This is illustrated in the lower panel of Fig. 3.

When an unpolarised beam moves through the oscillating magnetic field like shown in Fig. 4, then the atoms have different options to reach the HFS α1 from every other substate. Thus, the interference of the possible transitions determines how many atoms will end up in α1. At the optimal conditions most atoms within the $F=1$ multiplet will be in this state after the magnets. Thus, when at a certain magnetic field $B$ about 70% of the atoms are in the α1 state ($|m_J=+1/2, m_I=+1/2>$) and about 25% in β4 ($|m_J=-1/2, m_I=+1/2>$), the nuclear polarisation of the beam can approach $P\sim0.9$.

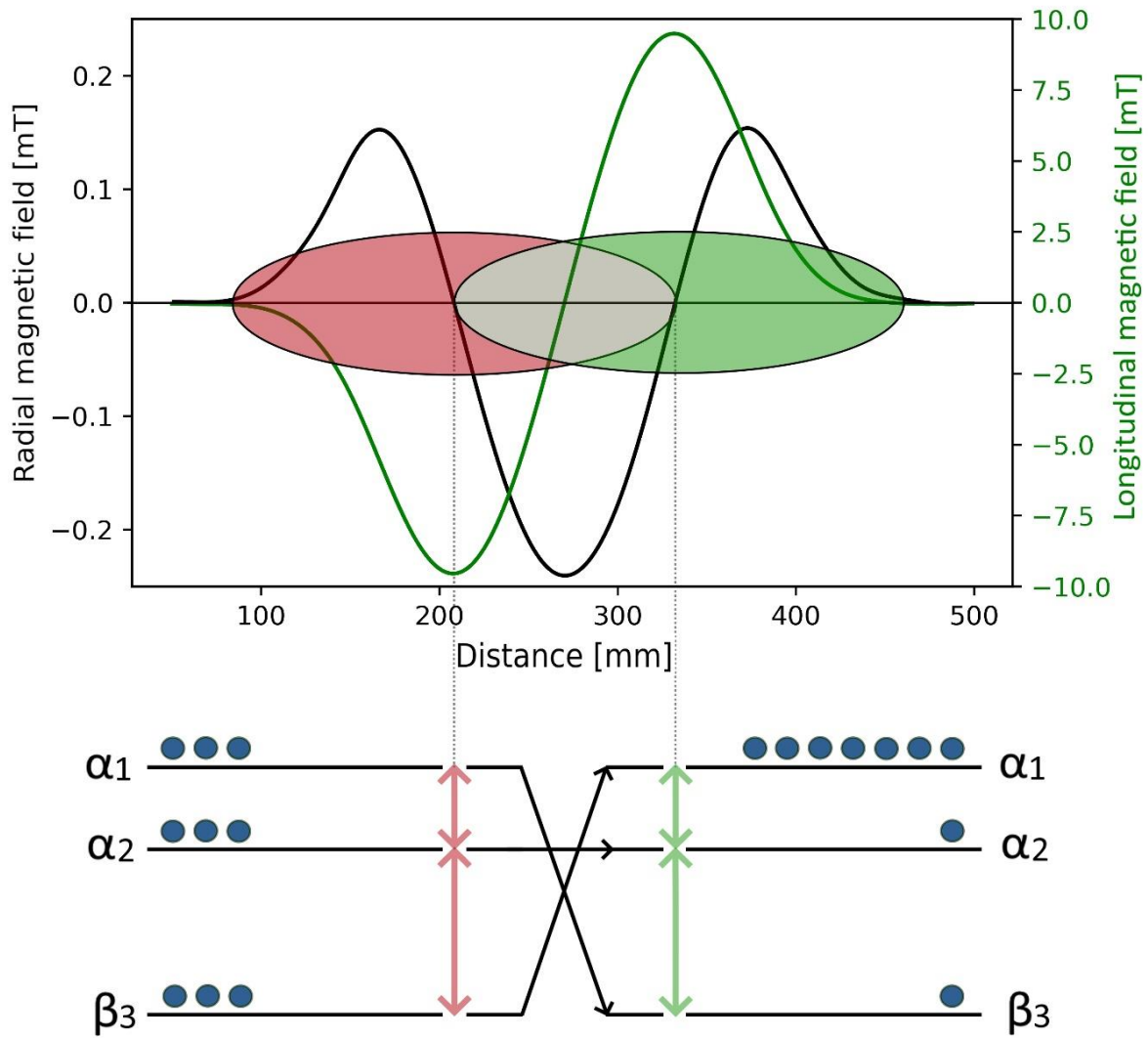

**Figure 4:** The induced oscillations of the radial magnetic field (black) along the two solenoids induce transitions within the hyperfine substates of the $F=1$ multiplet that can interfere with each other. By that the occupation numbers of the single substates can be pumped into just one state.

In other words: The static magnetic field of two opposing coils corresponds to a single radio-wave pulse with coherent and monochromatic photons, which induce radio-wave pumping between the states. The number of photons is determined by the amplitude of the radial magnetic field and exceeds easily the number of particles passing through by orders of magnitude. Thereby, the quantum mechanical interference of the transitions and the free choice of the magnetic field $B$ allow one to pump most of the metastable hydrogen atoms within the $F=1$ multiplet into a single hyperfine substate.

Our method is universal in various ways:
In principle, every beam particle can be polarised, if a hyperfine structure exists with $F \geq 1$, since at least 3 substates are required to induce the quantum interferences between the states. Similar to the examples shown above simulations and measurements have been made for metastable deuterium ($F=3/2$) and again a good agreement is observed. For ground state hydrogen and deuterium atoms simulations are available and a measurement with a deuterium beam at several 10 keV is in preparation. In this case the nuclear polarisation can be measured with a nuclear reaction polarimeter based on the known analysing powers of the $d+d \rightarrow t+p$ fusion reaction [21,22].

As shown in Ref. [23] the hyperfine structure of $^3$He$^+$ ions, apart from the opposite sign of the nuclear g-factor $g_{3He^+}= -4.2550996069$, is very similar to that of the hydrogen atom. Simulations for a 5 keV beam show that a beam polarisation of $P \sim -0.8$ at a magnetic field amplitude of about 1 mT is possible even for large intensities (see Fig. 5). Thus, polarised $^3$He$^+$ ion beams for stripping injection into storage rings could be produced, which can be used as an effective polarised neutron beam, because the nuclear spin is dominated by the neutron [24].

Beside the interaction of the nuclear and electron spins, also the smaller interaction of the nuclear spin and the rotational magnetic moment $J$ of a molecule, e.g. for $H_2$ or $D_2$, produce a hyperfine structure that can be manipulated in this way. This was already shown at photon energies of about $10^{-12}$ eV for $D_2$ molecules [25]. Another option can be the mutual interaction of the nuclear spins that can be found, e.g. in a beam of ortho-water. This might allow to polarise water by pumping the molecules into a single ortho-substate.

The same method should also work by sending corresponding pulses on a substrate. If a probe at rest is fixed inside a longitudinal and a radial set of two coils, the corresponding single radio waves can be induced by RF generators. We note that 1 W of induced RF power at a frequency of a few MHz corresponds to $10^{28}$ photons/s. Thus, even macroscopic amounts of material can be polarised and the nuclear polarisation be measured with NMR.

With this rather simple and cheap method many applications are within reach. Beside the example of a $^3$He$^+$ ion source the design of the coils can be adapted for different polarised ion sources, e.g. to investigate the spin dependence of nuclear reactions at particle accelerators [3]. In atomic physics precision measurements of the Breit-Rabi diagrams for different atoms, molecules and their ions are in range to test QED predictions [13], in principle even for anti-matter [25]. It is possible to control photon energies down to the peV range with feV uncertainties and below by using them like a laser pulse in laser spectroscopy. Another option would be the production of polarised tracers for medical applications or a new type of low-field NMR/MRI by polarising the probe just before the measurement.

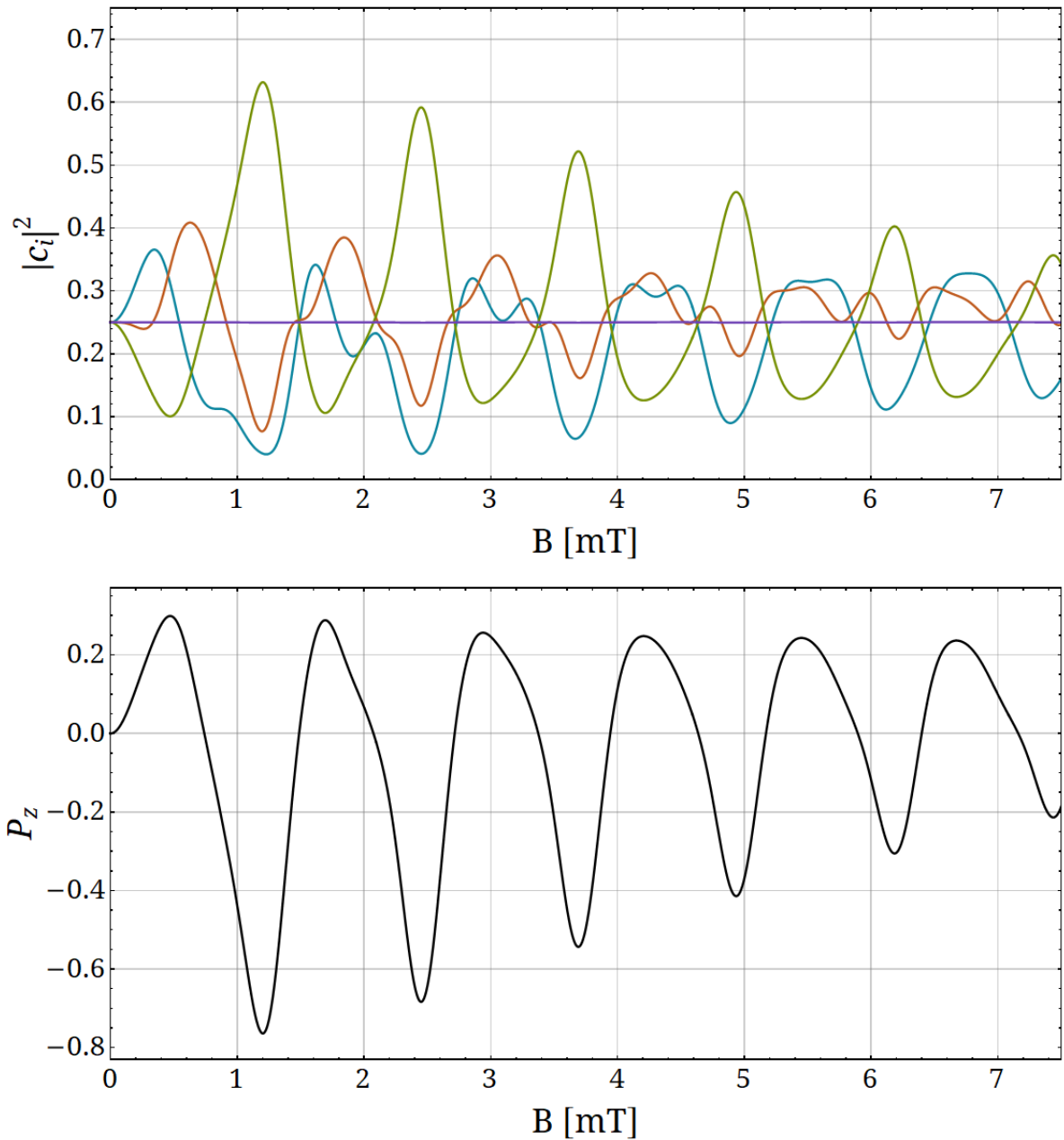

**Figure 5:** Simulated occupation numbers of the different hyperfine substates of $^3$He$^+$ ions at a beam energy $E_{kin}$= 5 keV passing the field coils with λ=10 cm (upper figure).
Nuclear polarisation $P_z$ of the $^3$He$^+$ ions in a strong magnetic field (Paschen-Back region) behind the solenoids (lower figure).

Another very attractive application is the production of nuclear polarised fuel to feed fusion reactors since the $d+t$, the $d+^3He$ [5] and the $p+^{11}B$ fusion reaction rates strongly depend on the orientation of the nuclear spins. The different reactor types are fed either with fast neutral atomic beams or frozen pellets of deuterium, tritium or HD molecules. Our simulations show that the production of polarised intense hydrogen, deuterium and tritium beams at the energies of several 10 up to 200 keV are possible and will be tested for deuterium in a coming experiment. Beside pellets filled with polarised $^3$He gas, it should be possible to neutralise a polarised $^3$He$^+$ ion beam by charge exchange with alkali metal vapour in a strong magnetic field to decouple electron and nuclear spins during the electron capture. To produce deuterium pellets several options can be investigated. Either polarised deuterium atoms are recombined into polarised molecules [27] or polarised $D_2$ molecules in a thermal beam are directly polarised due to their $I\leftrightarrow J$ interaction and then collected by freezing them as polarised ice in a strong magnetic field. With this method the production of polarised fuel is possible to test polarisation effects in fusion plasmas, both for magnetic confinement [28] and laser-induced fusion. Together with the recent confirmation of polarisation conservation in a laser-induced $^3He$ plasma [29], this paves the way for 'polarised fusion' energy production.

## Methods:

Experimental setup:

This simulation was verified with an experimental setup shown in Fig. 1 similar to Ref. [1]: With an ECR source, fed with hydrogen gas, a beam of protons and $H_2^+$ ions is accelerated to energies between 0.5 and 3 keV. The molecular ions are separated by a Wien filter, which in parallel helps to decrease the velocity distribution of the protons before they reach a caesium vapour cell. By charge exchange with the caesium unpolarised metastable hydrogen atoms in the $2S_{1/2}$ state are produced. Afterwards, a spinfilter [2] allows to control the population number of the substates. It filters out metastable atoms in a single hyperfine substate, either α1 or α2, both α substates equally populated or unpolarised metastable atoms can be passed through. At the same time its longitudinal magnetic field acts as a quantisation axis and is needed to align the spins. Next, the metastable atoms pass a pair of solenoids with opposite field directions to induce a single sinusoidal magnetic field oscillation. Another spinfilter can then filter the metastable atoms in the single α substates and quench all others into the ground state. The relative occupation numbers of these states are determined when the residual metastable atoms are quenched into the ground state by a strong electric field (Stark effect) and the induced Lyman-α photons are detected with a photomultiplier. Thus, the relative occupation numbers of the substate α1 and α2 are measured as function of the magnetic field in the centre of the Sona coils. The observed occupation numbers of the single HFS can be compared with the simulations discussed above.

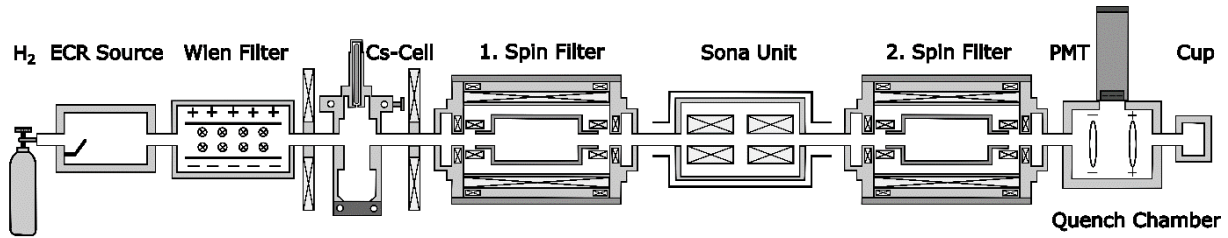

**Figure 1:** The experimental setup.

When the first spinfilter defines as polariser the occupation numbers of the single hyperfine substates, e.g. only atoms in the substate α1 are allowed to pass through, then the second spin filter works as an analyser. The only parameter to change during the measurements is the current through the Sona coils to manipulate the magnetic field amplitude, i.e. to ramp the magnetic field *B*. The photomultiplier signal is directly proportional to the number of metastable atoms in the single substate and can be plotted as function of the current through the coils.

Theory:

The formalism to calculate the occupation numbers of the hyperfine substates with perturbation theory is introduced in Ref. [3] and [4].

Without an external magnetic field, the hydrogen atom is well described by the total angular momentum $\vec{F} = \mathbb{1} \otimes \vec{J} + \vec{I} \otimes \mathbb{1}$ with $\vec{J}$ being the total angular momentum of the electron and $\vec{I}$ the spin of the nucleus. Therefore, the eigenbasis of the hydrogen system including fine structure and hyperfine splitting is given by $|F, m_F\rangle$. The additional time dependent perturbation enters in the experiment by the external magnetic field. Subsequently, the Hamiltonian describing the hyperfine splitting serves as the unperturbed one as the eigenproblem is known,

$$H_{HFS}|F, m_F\rangle = E_F|F, m_F\rangle.$$

The time evolution is controlled by the Schrödinger equation. At all times the wave function can be expressed as a linear combination of the unperturbed eigenbasis

$$|\psi(t)\rangle = \sum c_i(t) |F_i, m_{F_i}\rangle.$$

This leads to a system of coupled differential equations given by

$$i\hbar \dot{c}_k(t) = \sum c_i(t) e^{\frac{-i(E_i - E_k)t}{\hbar}} \langle k|V(t)|i\rangle.$$

The time dependent perturbation potential is given by the external magnetic field $B(t)$ interacting with the different magnetic moments of the hydrogen atom shown in the following expression

$$V(t) = \left(g_J \mu_B \frac{\vec{J}}{\hbar} - g_I \mu_k \frac{\vec{I}}{\hbar}\right) \cdot \vec{B}(t),$$

that has still much less influence than the hyperfine splitting itself.

As the energy gap for small magnetic fields, i.e. $\mu_B |B|/\Delta E \ll 1$, to the next neighbouring set of states is far enough for the metastable $2S_{1/2}$ set, only these four states need to be taken into account for the calculations. This gives us the following set of coupled differential equations

$$i\hbar \dot{c}_{\beta_4} = c_{\beta_3} e^{-i\Delta E_{HFS} t/\hbar} \frac{B_r(g_J\mu_B + g_I\mu_k)}{2\sqrt{2}} + c_{\alpha_2} e^{-i\Delta E_{HFS} t/\hbar} \frac{B_z(g_J\mu_B + g_I\mu_k)}{2}$$
$$- c_{\alpha_1} e^{-i\Delta E_{HFS} t/\hbar} \frac{B_r(g_J\mu_B + g_I\mu_k)}{2\sqrt{2}}$$

$$i\hbar \dot{c}_{\beta_3} = c_{\beta_4} e^{i\Delta E_{HFS} t/\hbar} \frac{B_r(g_J\mu_B + g_I\mu_k)}{2\sqrt{2}} - c_{\beta_3} \frac{B_z(g_J\mu_B - g_I\mu_k)}{2}$$
$$+ c_{\alpha_2} \frac{B_r(g_J\mu_B - g_I\mu_k)}{2\sqrt{2}}$$

$$i\hbar \dot{c}_{\alpha_2} = c_{\beta_4} e^{i\Delta E_{HFS} t/\hbar} \frac{B_z(g_J\mu_B + g_I\mu_k)}{2} + c_{\beta_3} \frac{B_r(g_J\mu_B - g_I\mu_k)}{2\sqrt{2}}$$
$$+ c_{\alpha_1} \frac{B_r(g_J\mu_B - g_I\mu_k)}{2\sqrt{2}}$$

$$i\hbar \dot{c}_{\alpha_1} = -c_{\beta_4} e^{i\Delta E_{HFS} t/\hbar} \frac{B_r(g_J \mu_B + g_I \mu_k)}{2\sqrt{2}} + c_{\alpha_2} \frac{B_r(g_J \mu_B - g_I \mu_k)}{2\sqrt{2}}$$
$$+ c_{\alpha_1} \frac{B_z(g_J \mu_B - g_I \mu_k)}{2} \quad ,$$

where $\Delta E_{HFS}$ is the hyperfine-structure splitting energy with $\Delta E_{HFS}(1S_{1/2}) = 5.87$ µeV for the ground state 1S and $\Delta E_{HFS}(2S_{1/2}) = 0.734$ µeV for the metastable state.

The formalism allows us to simulate the occupation numbers of the beam atoms during their flight through the magnetic coils that induce a static field. Solutions of the set of the coupled differential equations including an integration along a Gaussian beam profile are given in Fig. 2.

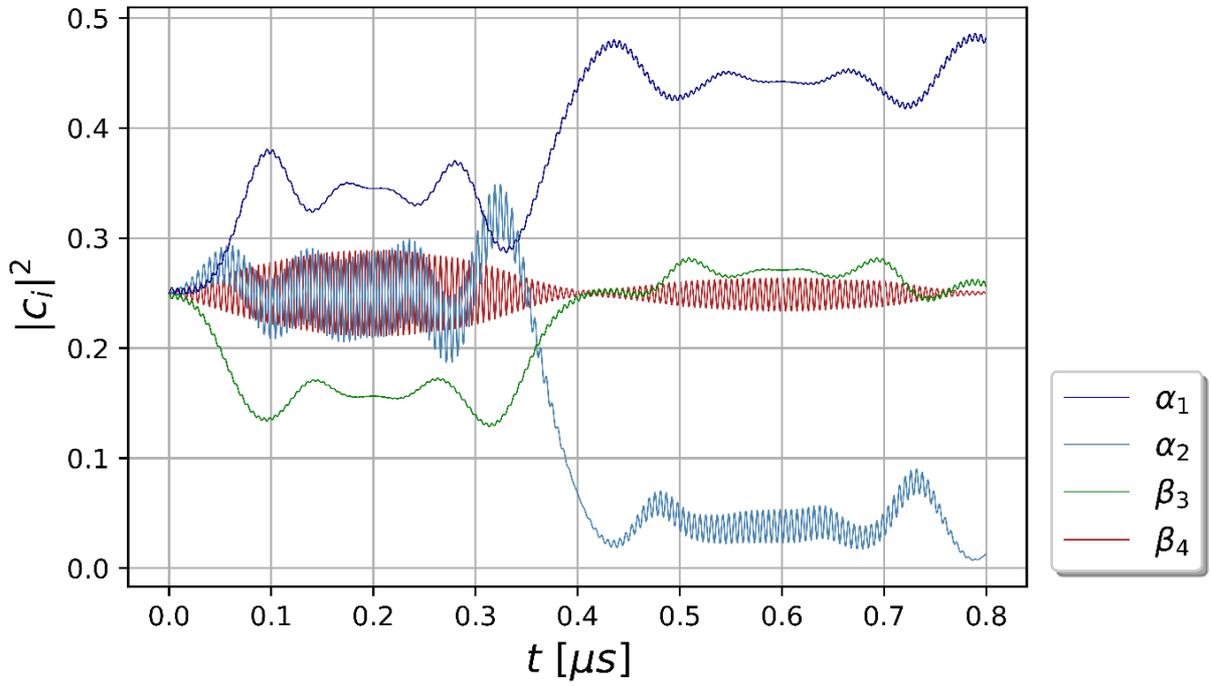

a.)

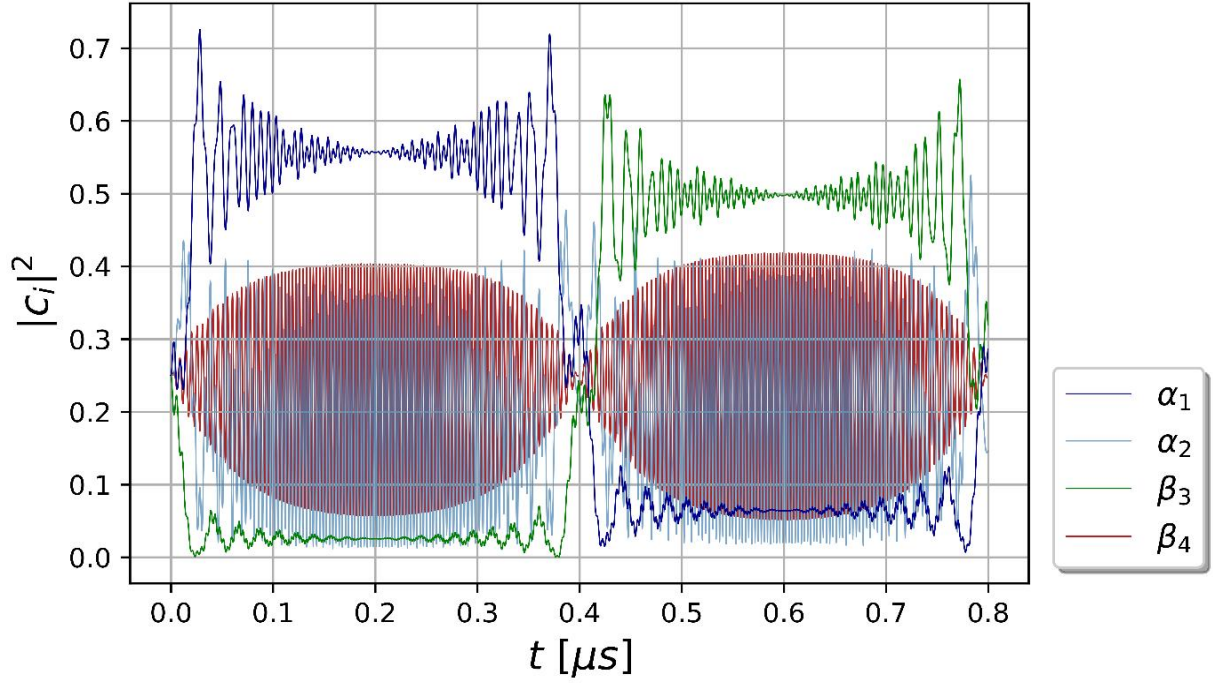

b.)

**Figure 2:** The time-resolved occupation numbers of the HFS during their flight through the Sona coils when an unpolarised beam of metastable hydrogen atoms moves through with a beam energy of 1 keV. In case (a.) the amplitude of the magnetic field B is 1 mT and for (b.) 10 mT.

The simulations are produced with an ideal sinusoidal function for the magnetic field in beam direction and show the occupation numbers of the four different metastable states in time. The magnetic field amplitude used for Fig. 2a.) is $B = 1$ mT and for the second $B = 10$ mT. The corresponding wavelength of the magnetic field is set to $\lambda = 0.35$ m and the beam energy is fixed to $E_{\text{Beam}} = 1$ keV. The interaction of the entire magnetic field then brings the occupation numbers of the atoms in an unpolarised beam into oscillations and behind the solenoids these occupation numbers are not equal any more, i.e. the beam is polarised.

At the beginning the radial magnetic field leads to a rotation of the electron spin due to the Larmor precession around the radial field that induces first a transition between the states α1↔α2 and α2↔β3. In the centre of the first coil these transitions are frozen due to the absence of a radial field, but now the hyperfine beat [5] between the states α2↔β4, induced by the longitudinal magnetic field component $B_z$ in the couple differential equations, is visible. When the radial component is changing again and becomes negative, the first oscillations are inversed and after the first coil the beam would be again unpolarised, if there would be no zero-crossing between the coils. This crossing exchanges the direction of the quantization axis and corresponds to a phase shift of the oscillations observed before. By that the occupation numbers at the end of the two opposing coils are unequal and the beam can get polarised. Depending on the magnetic field amplitude of the longitudinal magnetic field the radial component and, therefore, the Larmor frequency are modified which can change the occupation numbers at the zero-crossing in the centre and by that the polarisation of the beam at the end.

For different starting conditions, i.e. only substate α1 is occupied ($|c_{\alpha 1}|^2$=1), full electron polarisation ($|c_{\alpha 1}|^2$=0.5=$|c_{\alpha 2}|^2$), and an unpolarised beam ($|c_i|^2$=0.25) these simulations show significant differences. This is demonstrated in Fig. 3 showing the occupation numbers for the substate α1 plotted as function of the magnetic field $B$.

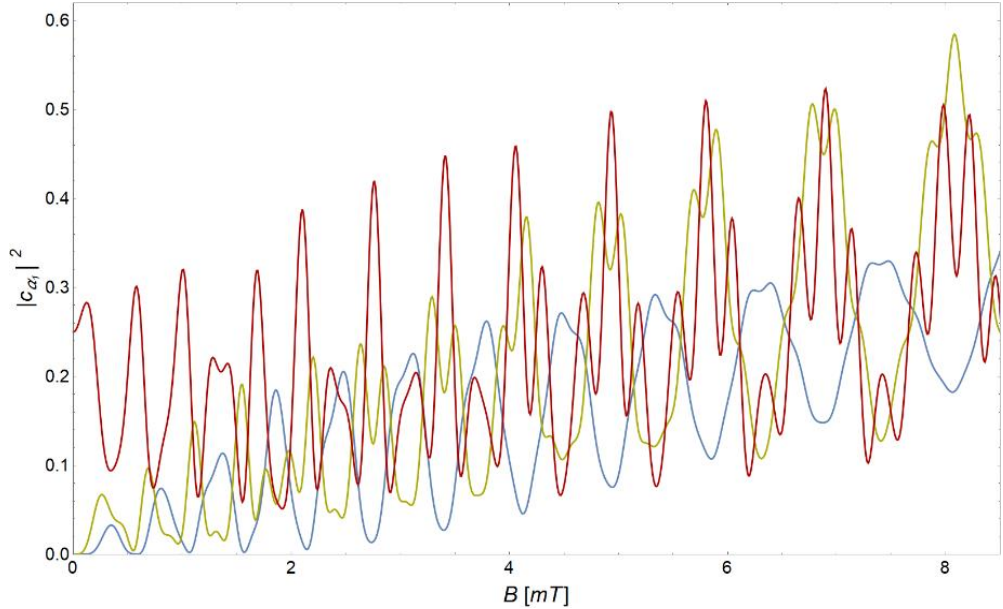

**Figure 3:** The calculated relative occupation numbers of atoms in the HFS α1 when a beam of metastable atoms at 1.5 keV is passing through the magnetic field shown in Fig. 1 of the main text. The oscillations depend strongly on the starting conditions that are $|c_{\alpha 1}|^2$=1 (blue), $|c_{\alpha 1}|^2$=0.5=$|c_{\alpha 2}|^2$ (green), and $|c_i|^2$=0.25 (red) for an unpolarised beam.

Especially the interference of the transitions can be constructive or destructive for the occupation numbers, which can be explained best is the photon model (see Fig. 4 in main text):

- If at the beginning only the substate α1 is populated these atoms will be transferred into the state β3 until no photons are absorbed. Thus, when the photon absorption induces the transition α1↔α2 the atoms in the substate α2 survive the zero-crossing of the magnetic field unchanged and can go back into α1 by absorbing a second photon.
  Then, the state α1 can be found again and will be populated at maximum when this transition is fitting at its best. Here, the α1↔α2 transition is constructive, but the superposition with the even possible α2↔β3 transition is destructive, because atoms can get lost into β3 before they reach α1 again.

- When both α substates are populated equally the α1↔α2 transition does not play a role at the beginning. Now the α2↔β3 transition will populate the β3 state and from here the atoms are transferred into the α1 substate due to the Sona transition. Thus, now the α2↔β3 transition is constructive and the α1↔α2 transition will be destructive, because in the second part of the Sona transition atoms now can get lost from α1 into the less occupied substate α2. These phase shifts can be clearly seen in Fig. 3.

- If all HFS are equally populated the situation is much more complicated. Now an interference of the different options to populate the substate α1 again will happen.

Nevertheless, from Fig. 3 it is obvious that the α1↔α2 transition is still destructive, because there is no phase shift compared to case b. But the phase shift for the α2↔β3 transition even shows that this transition is now destructive too.

Systematic effects:

An important aspect of the understanding of this new polarising method are the different sources of systematic effects that determine the half width of the resonances in the measured spectra.

- In comparison with other spectroscopy measurements is the interaction time of the photon with the atom $\Delta t$ ~1 µs rather long due to the relatively small velocity of the beam $v \ll c$. Thus, the Heisenberg uncertainty relation

$$\Delta E \cdot \Delta t \sim h$$

defines directly the minimum for the full half width of the resonances $\Delta E$ that follows directly as

$$\Delta E \sim h/\Delta t = h \cdot f_0 = E_{Photon} \quad.$$

Thus, the distance between the peaks is twice the photon energy when the magnetic field axis is recalibrated due to the known Breit-Rabi diagram into an energy scale, and the minimum full half width of the resonances is equal to one photon energy. This is even obvious in the measured and simulated spectra.

- The velocity distribution of the beam atoms (with the half width $\Delta v$) induces an uncertainty of the photon energy. By use of the Wien filter $\Delta v/v \leq 10^{-2}$ is reached and, therefore, $\Delta E_{Photon}/E_{Photon} \leq 10^{-2}$. But this effect is accumulating: For example, for the $n^{th}$ peak in total (2n-1) photons must be absorbed and the uncertainty will be
$\Delta E_n = E_{Photon} + (\sqrt{(2n-1)} \cdot \Delta v/v \cdot E_{Photon}) = E_{Photon} \cdot (1+\sqrt{(2n-1)} \cdot \Delta v/v)$.

- Inhomogeneities of the magnetic field have a similar impact to the measured spectra. Only for an infinitely long solenoid the magnetic field inside does not depend on the axial distance ρ. The shorter the coils are, the stronger the increase of $B_{max}$ and, therefore, of B as function of r will be. With a beam intensity distribution along *r*, different parts of the beam will experience slightly different B with $\Delta B/B$ = const. The beam profile is not touched when the current through the solenoids is increased. Therefore, the magnetic field is smeared and the resonances are broadened. Of course, this effect increases linear with the magnetic field. To minimize this effect, the solenoids should be longer or the inner diameter must be decreased. The amplitude of this effect can be investigated directly when the magnetic field axis is shifted relative to the beam axis. Thus, the increase of B(r) can be calculated and measured. Of course, this effect gets stronger with larger offsets.

- The number of photons that can be absorbed by the particles depends on the amplitude of the radial magnetic field component. Therefore, the beam profile is influencing the measured spectra, because when more atoms are found at a larger radial distance $r$ even the probability of a transition becomes larger.

Due to apertures in the beam line the maximal beam diameter is limited to about 20 mm. If now a Gaussian beam profile is assumed that is described due to its root mean square deviation $0 \leq \sigma \leq \infty$, the corresponding spectra can be simulated via an integration of the spectra for different $r$. For a perfect beam on the magnetic field axis ($\sigma = 0$) no transitions are observed, because the radial magnetic field component will be zero. When an unpolarised beam of metastable hydrogen atoms is entering the solenoids, the simulated spectra for the occupation number of the hyperfine substate α1 are shown in Fig. 4 for σ=5 mm and $\sigma = \infty$, i.e. a homogeneous beam distribution along $r$. Still, all resonances appear at the same magnetic fields, but the amplitudes of the single transitions are changing due to the different probabilities to absorb a photon.

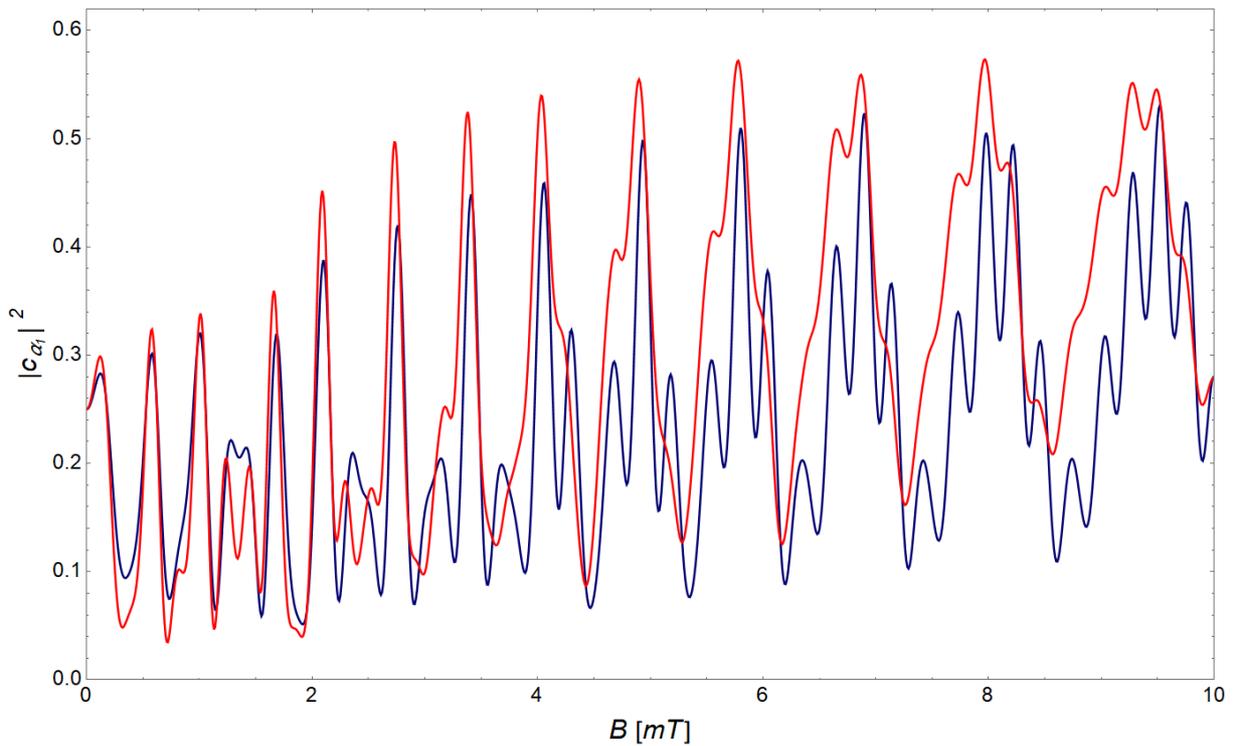

**Figure 4:** Relative occupation numbers of the HFS α1 for a beam of unpolarised metastable hydrogen atoms at a beam energy of 1.5 keV with a Gaussian beam distribution with $\sigma = 5$ mm (blue) or $\sigma = \infty$ (red) entering the solenoids.

These effects are the reasons for the increasing minima between the resonances due the slightly increasing half widths (see Fig. 2a of the main text), and the degrading resolution of the peaks (see Fig. 2b). More homogeneous magnetic fields and a better velocity filter will decrease this effect, but might decrease the number of measured photons in the photomultiplier.

Acknowledgments

This work has been supported by the ATHENA consortium (Accelerator Technology HElmholtz iNfrAstructure) in the ARD programme (Accelerator Research and Development) of the Helmholtz Association of German Research Centres. The authors want to thank Sahil Aswani for the preparation of the sinusoidal coils and Berthold Klimczok for the technical design of several components.

Authors contributions

R.E. prepared and performed the experiment, guided the data analysis and wrote the manuscript. T.E-K. made simulations and carried out data analysis. N.F. and C.H. made simulations and wrote parts of the manuscript. S.P., V.V., M.W. and J.W. performed measurements. L.K. and N.H. performed measurements and carried out data analysis. C.K. performed the experiment, made simulations and carried out data analysis. H.S. designed the magnetic fields. T.S. helps to setup the experiment. M.B. organized funding, contributed to analysis and wrote parts of the manuscript.